\documentclass[conference]{IEEEtran}

\usepackage{balance}

\IEEEoverridecommandlockouts 
\usepackage[dvips]{graphicx}
\usepackage{listings}
\usepackage[OT4,T1]{fontenc}
\usepackage[cmex10]{amsmath}
\usepackage{url}
\usepackage{multirow}

\usepackage{multirow}
\usepackage{dblfloatfix}
\usepackage{siunitx}
\usepackage{booktabs}
\usepackage{longtable}
\usepackage{multicol}

\usepackage{cite}
\usepackage{amsmath,amssymb,amsfonts}
\usepackage{graphicx}
\usepackage{textcomp}
\usepackage{xcolor}
\usepackage{hyperref}

\usepackage{multirow}
\usepackage{rotating}

\usepackage{mdframed}
\usepackage{lipsum}
\usepackage{soul}

\usepackage[]{algorithm2e}

\title{Verification and Optimization of Cyber-Physical Systems: Preprint for FedCSIS}
\author{
\IEEEauthorblockN{Reza Soltani, Eun-Young Kang, Juan Esteban Heredia Mena}
\IEEEauthorblockA{
University of Southern Denmark\\
The Mærsk Mc-Kinney Møller Institute\\
SDU Software Engineering, Odense, Denmark\\
Email: \{resol, eyk, jehm\}@mmmi.sdu.dk}
}

\begin{document}
\maketitle              

\begin{abstract}
Optimizing CPS behavior in terms of energy consumption can have a significant impact on system reliability. The environment influences the system's behavior, and neglecting the environmental behavior has an indirect negative impact on optimizing the system's behavior. In this work, to increase the system's flexibility, the behavior of the environment is modeled dynamically to apply the disorderliness of its behavior. The resulting models are formally verified.  By examining the past environmental behavior and predicting its future behavior, energy optimization is done more dynamically. The verification results acquired using a UPPAAL-SMC show that the optimization of system behavior by predicting the environmental behavior has been successful. Our approach is demonstrated using a case study within an I4 setting. 
\end{abstract}

\section{Introduction} \label{Introduction}
\IEEEoverridecommandlockouts\IEEEPARstart{C}{yber-physical systems (CPS)} are continuously developing and have become an integral part of Industry 4.0 (I4). The I4 revolution relies on the interconnectivity of machines and automation of processes to improve factory productivity. Modern industry uses assets designed to do different tasks; for example, a robotic arm can assemble, palletize, and solder. Consequently, many production cells can perform the same activity, but each cell's production time differs. For instance, total automation, sustainable production, and efficient scheduling are challenges that need to be addressed to achieve the I4 objective.    

According to the United Nations Goals, production sustainability is one of the goals for industry \cite{unitednations_2015}. One of the key factors in CPS is energy consumption, and neglecting energy consumption limits the reliability and safety of such systems. We differentiate between two types of techniques for reducing the energy consumption of a system: whether a CPS has to reconstruct its behavior or make structural modifications such as using light materials, different types of batteries, and efficient motors. Structural modifications can reduce energy consumption but are costly; instead, behavioral changes modify the device software and are cheap.

One of the challenges in CPS analysis is predicting the environmental behavior and modeling the environment since its behavior may change depending on various factors. Although CPS is highly affected by the environment, it has no control over it. For example, in an autonomous vehicle system, the system's performance is based entirely on the analysis of environmental behavior. Knowing information such as how often we are likely to see each sign or which sign is less likely to be seen based on the current situation allows the system to make optimal decisions. For example, the speed can increase with a gentler slope if it is likely to see a stop sign. Sometimes no strong logic can be found for modeling the environment. In this regard, the challenge is run-time data collection, and the system's ability to dynamically model the environment based on these data increases the system's flexibility. One way to do this is to use learning algorithms to study the past and predict future environmental behavior.

In this paper, energy-aware timed behaviors of CPS are specified in Stochastic Hybrid Automata (SHA) \cite{SHA}. By considering the cost of each behavior, the amount of energy consumption of the system for different modes is obtained. In this way, The minimum and maximum energy consumption of the system can be calculated. Due to the dependence of energy consumption on environmental behavior and the disorderliness of this behavior, the environment with which CPS is associated is dynamically modeled. 

Our earlier works \cite{ProbabilisticVerification, energyconstraints, VerificationCPS, Formalverification} verified the safety of CPS (including the controller and physical parts) and analyzed performance in terms of time and resource constraints: In order to model the unpredictable behaviors of environment,  fixed probabilities were considered and allocated to the possible transitions. Whereas our current work utilizes dynamic probabilities instead of the predetermined probability. The main contributions of this paper are:
\begin{itemize}
    \item Formal modeling of energy-aware timed behaviors of CPS in SHA that captures discrete and continuous behaviors of both the controller and its environment and verifying the system using a statistical model checker, UPPAAL-SMC.  
    
    \item Dynamic modeling of the environment to increase the flexibility of the system against the disorderliness behavior of the environment. 
    
    \item Optimize the behavior of the system in terms of energy consumption by predicting the future environmental behavior.  
\end{itemize}

In this work, we use the dynamic probability for each environmental incident automatically updated after a certain number of iterations. According to the past behavior of the environment, the system can predict its future behavior and make decisions based on it. Thus, the environmental behavior of the system is modeled more dynamically.

The rest of the paper is organized as follows. Section \ref{Methodology} presents our methodology. The industrial case study we used to demonstrate the applicability of our framework is presented in Section \ref{Case Study}. Section \ref{model} shows the modeling and verification results. Section \ref{Related work} provides related works, and Section \ref{future works} describes research challenges and intended future works.

\section{Methodology} \label{Methodology}

\subsection{Energy Optimization}

The behavior of CPS can be optimized by improving the behavior of its components without structural changes, such as energy modes. We conduct optimization by addressing energy consumption; We model the continuous behavior of the system through SHA, which enables us to find the best path of having a specific behavior with minimum energy consumption. By using statistical analysis, the probability of energy optimization has been investigated up to a certain confidence degree. The behavior of the environment, which directly impacts the energy consumption process, has also been studied.

\subsection{Modeling environmental behavior}
To model and predict the environmental behavior of CPS, it is essential to learn from its previous behavior. As discussed in Section I, the dynamic probability is used because the behavior of the environment can change, and modeling it with fixed probabilities will significantly simplify and limit the system model. Therefore, the current work uses variable probabilities for each environmental incident by considering its behavioral disorderliness. These probabilities are updated in each period using closed-loop control and by analyzing the output of the system. To do this, the occurrence time of each event is stored in an array of integers. After a certain number of iterations, this array is sorted using the ascending sort algorithm, and the number of occurrences of each event is calculated. Based on that, the probability of its recurrence in the future is predicted. This possibility will be updated automatically on each iteration. Thus, instead of using fixed and predetermined probabilities for each event, by considering the system's past behavior (environment) and predicting future behavior based on it, dynamic probabilities updated in each iteration are used.

\section{Case Study} \label{Case Study}
\begin{figure}[h!]
\centering
\includegraphics[width=0.50\textwidth]{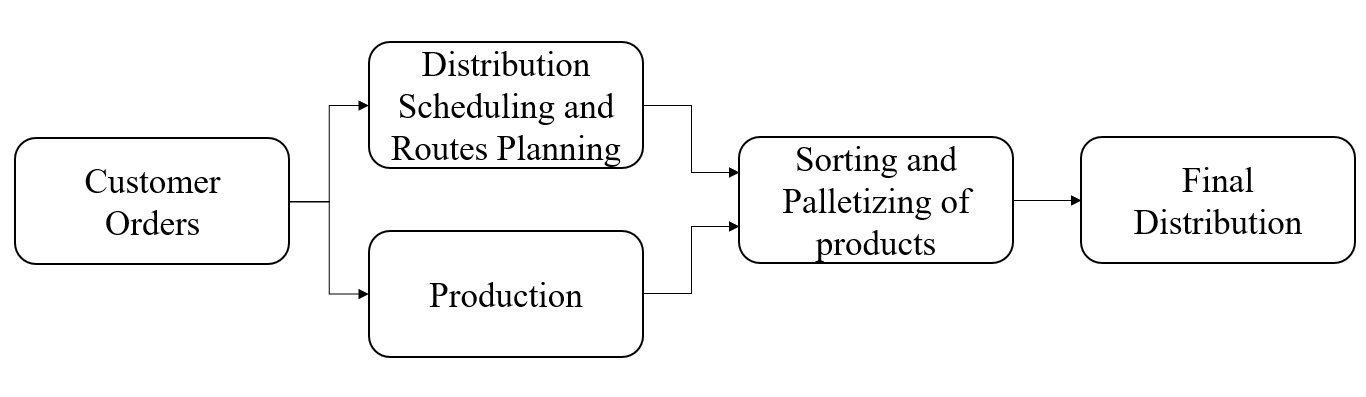}
\caption{Reduced System Architecture of a Factory}
\label{fig:SystemArchitecture}
\end{figure}

 In this section, we present a case of study that is part of a production plant. Fig. \ref{fig:SystemArchitecture} shows a reduced system architecture of a factory. In non-automatized factories, operators manufacture and package the final products. In parallel, the distribution schedule and routes of distribution are defined. In the next stage, operators sort elements for distribution via a truck.
 
The Industry 4.0 Laboratory at the University of Southern Denmark (SDU) \footnote{https://www.sdu.dk/en/forskning/i40lab} is working in a production line for drone assembling \cite{Lab40}. The current state of the project does not consider the distribution of the drones. We aim to add the capability of distribution to the production plant. The distribution system selects the elements from a storing warehouse, classifies and orders them. The expected final result is a pile of drone boxes ready to be picked by a truck. 

\begin{figure}[h!]
\centering
\includegraphics[width=0.30\textwidth]{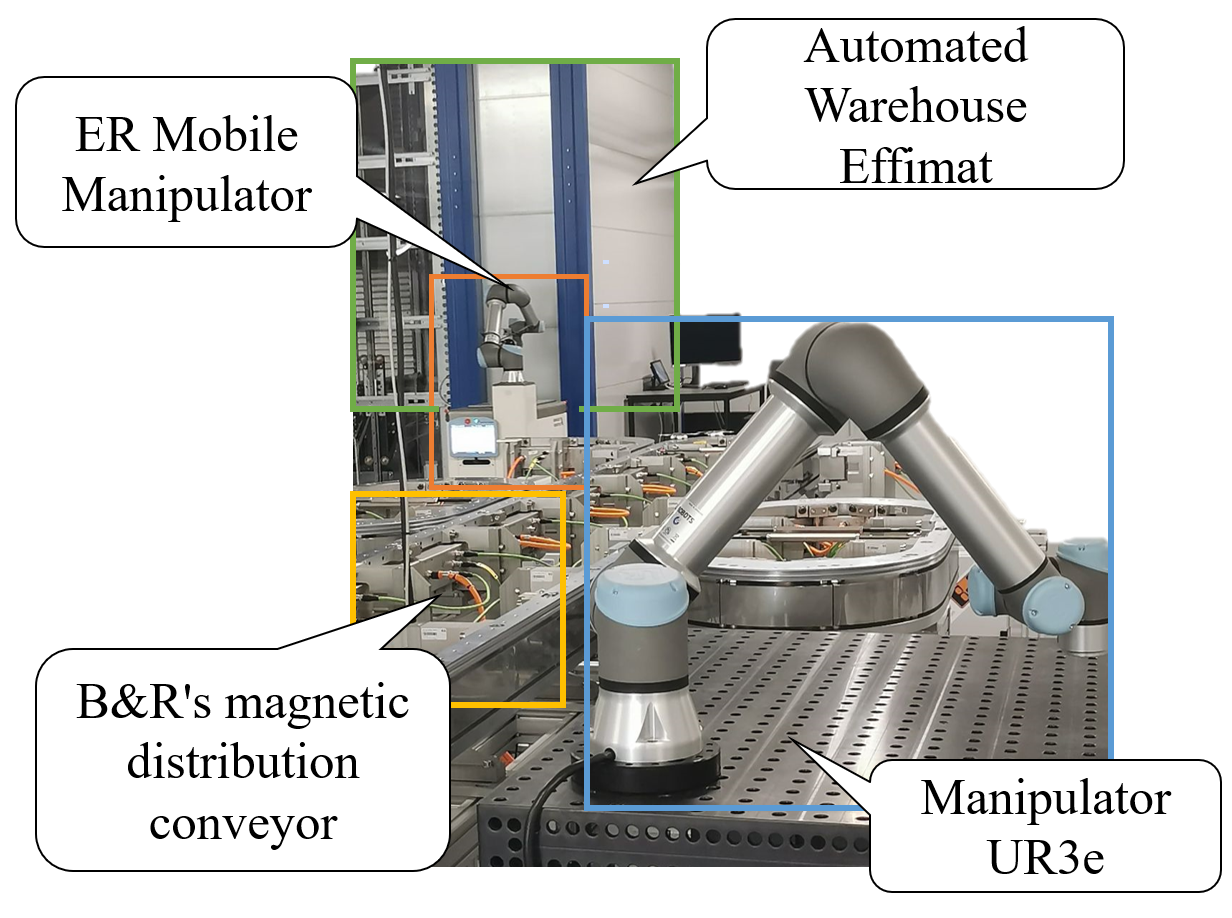}
\caption{Distribution system and its components}
\label{fig:Distributionsystem}
\end{figure}

In the proposed distribution system, the drones are obtained from the automated warehouse Effimat \cite{effiadmin_2021}. An ER mobile manipulator \cite{enabledrobotics} picks the products from the warehouse and places them in the B\&R's magnetic distribution conveyor. A robotic manipulator UR3e \cite{collaborativeroboticautomation} sortes and palletizes the packages using distribution schedule and route plan. Fig. \ref{fig:Distributionsystem} shows the elements that compose the distribution system.

\begin{figure}[h!]
\centering
\includegraphics[width=0.25\textwidth]{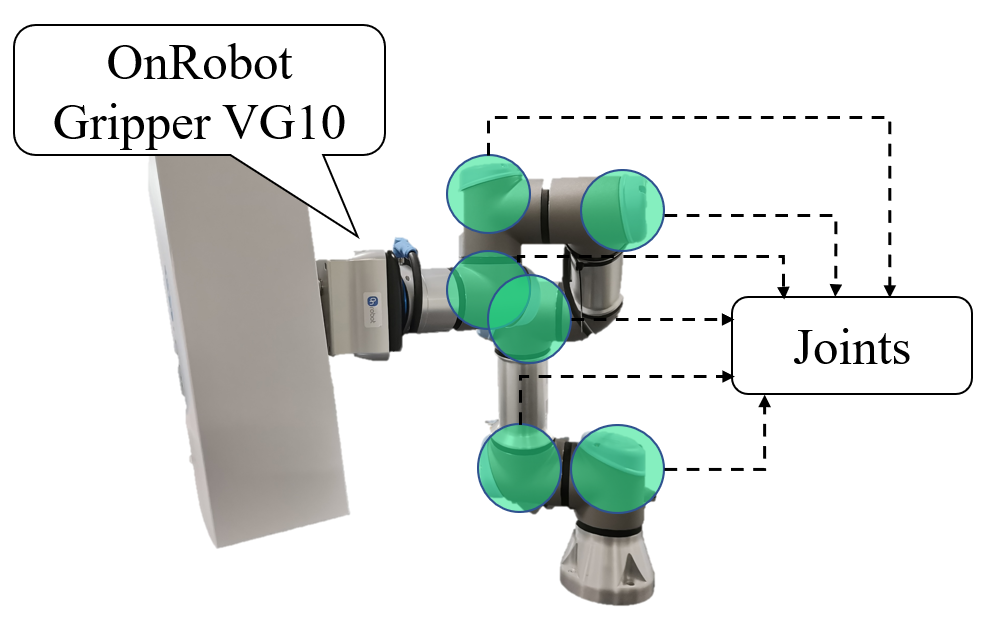}
\caption{Robot UR3e joints and griper VG10}
\label{fig:UR3e}
\end{figure}

We particularly focus on the sorting capability, which is done by the manipulator UR3e. The manipulator UR3e is an anthropomorphic robotic arm made by Universal Robots \cite{collaborativeroboticautomation}. The robot UR3e possesses six axes of rotation, as is shown in Fig. \ref{fig:UR3e}. 

In the sorting activity, the robot picks the objects from position \textit{A} and then places the package in position \textit{B} or \textit{C} according to the distribution plan. The frequency of packages depends on the customer orders and the distribution plan. The system includes a position idle. When the robot is in the idle position, it consumes less energy than in the other positions. 
%
One of the objectives of this study is to design an energy optimal robot behavior based on environmental behavior. Therefore, we measure the energy consumption of each of the robot movements and the rate of consumption of the robot in a static position, which is presented in Table \ref{energyconsumption}. 

\begin{table}[h]
\centering
\caption{Energy consumption of each of the robot movements}
\begin{tabular}{@{} c  c @{}}
\toprule
 \textbf{Movement /\ position} & \textbf{Energy Consumption} \\ \midrule
\textbf{A to B} & 374.0 [J]\\
\textbf{A to C} & 293.3 [J]\\
\textbf{B to A} & 196.1 [J]\\
\textbf{C to A} & 125.7 [J]\\
\textbf{B to Idle} & 198.6 [J]\\
\textbf{C to Idle} & 145.0 [J]\\
\textbf{Idle to A} & 112.3 [J]\\
\textbf{A } & 40.20 [J/s]\\
\textbf{B} & 39.87 [J/s]\\
\textbf{C} & 40.45 [J/s]\\
\textbf{Idle} & 37.45 [J/s]\\
\bottomrule
\end{tabular}
\label{energyconsumption}
\end{table}

\section{Modeling and verification results} \label{Model}

The described case study in Section III, is modeled in SHA, including the energy-aware timed transitions and the amount of energy consumed in each state. The operational semantics of both the system and its environment are formally specified in SHA, and the models are verified against given requirements through UPPAAL-SMC \cite{SMC} The verification results are displayed in Table II and discussed in Section \ref{Vresult}

\subsection{Modeling energy-aware behavior and environmental behavior prediction}

\begin{figure}[h!]
\centering
\includegraphics[width=8cm, height=14cm]{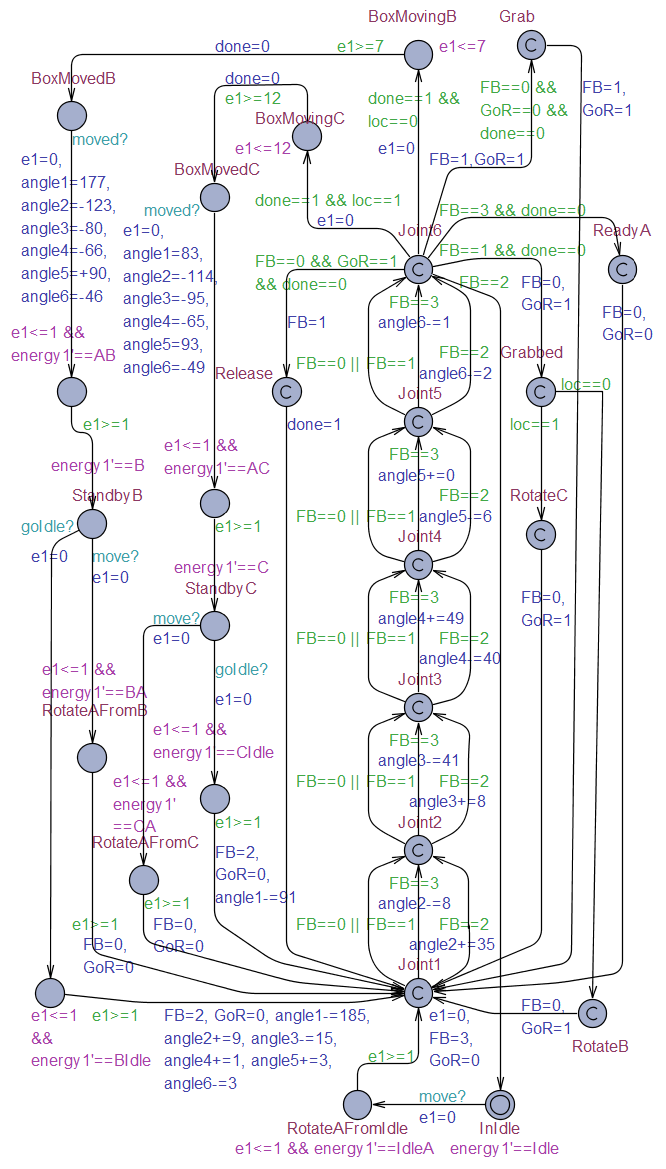}
\caption{Robotic arm automaton}
\label{fig:Behavior}
\end{figure}

Fig. \ref{fig:Behavior} shows the robotic arm automaton. Initially, the robotic arm is in idle mode, and with the arrival of the package, it receives a message through the synchronization channel \textit{move?}. In this case, each of the six joints of the robotic arm has to change the angle of their position to put the arm in the ideal position for state A; then, it will be ready to grab the package.  Depending on where the package goes, it will reset its joints' angles again and move the package to the desired location.
In this model, the amount of energy consumption for each transmission is calculated discretely according to Table \ref{energyconsumption}. The robotic arm consumes energy per time unit when it is in one of the three different modes (Idle, StandbyB, StandbyC). The amount of energy consumption for each of the three modes is calculated continuously. For example, the energy consumption per time unit when the robotic arm is in idle mode is calculated using formula \textit{energy'==Idle}. 

In this SHA model, the variant \textit{FB} is the functional behavior of the system, which can have an integer value between 0 to 3, representing moving up, moving down, going idle, and coming back from idle behaviors, respectively. The variant \textit{GoR} is the grabber's behavior which can have an integer value of 0 or 1, representing grabbing the box or releasing it, respectively. After moving box from position \textit{A} to \textit{B} or \textit{C}, it is decided whether to stay in the same state (StandbyB or StandbyC in Fig. \ref{fig:Behavior}) or to go into the idle mode to save energy.

\begin{figure}[h!]
\centering
\includegraphics[width=0.4\textwidth]{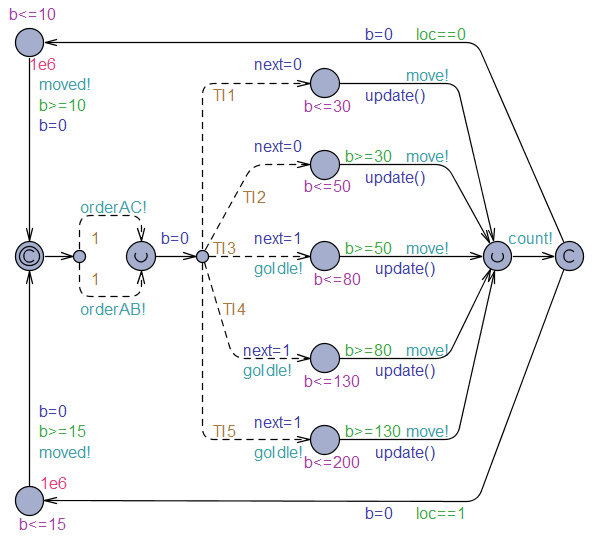}
\caption{Environment automaton}
\label{fig:Box}
\end{figure}

Such a decision is made with the help of the environment model. Fig. \ref{fig:Box} shows the SHA of the environment: 5 different periods are defined, which indicate the arrival time of the packages. The probability of occurrence of each is a variable of type integer, whose value is updated every 20 times the packages have arrived, and henceforth we call this period iteration.

The general goal of the environment SHA is: (1) to model the behavior of the environment dynamically to include its disorderliness behavior; (2) to change the behavior of the system based on the analysis of the past environmental behavior and to decide on the future behavior of the system based on a prediction. In the first case, five variables \textit{TI1}, \textit{TI2}, \textit{TI3}, \textit{TI4}, and \textit{TI5} are used for the probability of occurrence of each time interval as shown in Fig. \ref{fig:Box}. These variables are changing in each iteration, and this increases the possibility of including critical factors such as environmental disorderliness in modeling. In the second case, using the \textit{Update()} function, each time the boxes arrive is saved in the array, and every 20 times, at the end of the iteration, this SHA uses this function to decide on the behavior of the system for the next iteration. Algorithm \ref{algorithmm} shows how the \textit{Update()} function works.

In Algorithm \ref{algorithmm}, \textit{env} is an array of length 20 that saves the box arrival time. When the 20th box has arrived, it counts the number of times a box arrived in each time interval and updated the probabilities accordingly. Suppose the system goes into the idle mode and a box enters position \textit{A} earlier than a specific time. In that case, the system consumes more energy because the transition cost to enter to and exit from the idle mode is more than staying in standby mode. For this reason, this function makes a decision using the calculated probabilities for the next iteration. It should be noted that in some transitions, energy consumption may increase. Still, the purpose of this prediction is to determine whether, on average, a particular transition (going idle) will decrease or increase energy consumption. Verification results show that this algorithm has been successful in reducing energy consumption.

\subsection{Verification results} \label{Vresult}

In this section, we discuss the verification results that are given in Table \ref{Query}: \textit{energy} indicates the system's energy consumption in the presence of idle mode and \textit{energy2} indicates the system's energy consumption in the absence of idle mode.  The first five queries are related to energy consumption. As presented in the verification results, by predicting the behavior of the environment and changing the system's behavior accordingly, the amount of energy consumption is reduced by a probability of close to 100 (query 3). For example, in the table, in 3 hours (10800 seconds) and with an average of 20 runs, 60 kJ of energy consumption has decreased. In some transitions, there is a possibility that energy consumption in the case of having idle mode increases (queries 4 and 5).

\begin{algorithm}[]
\SetAlgoLined
\scriptsize
void update()\;
 
  \eIf{counter >= 19}{
    env[19]=b;\;
    
    TI1=0;\; TI2=0;\; TI3=0;\; TI4=0;\; TI5=0;\;
    
    counter=0;\; i= 0; j=1;
    
    \While{i < 20}{\;
    \While{j < 20}{\;
   \If{env[i] >= env[j]}{
   q1 =  sort[i];
    sort[i] = sort[j];
    sort[j] = q1;
   }
   j++;
    }
    i++;
   }
   i= 0;
   
   \While{i < 20}{\;
   \If{env[i] <= threshold1}{
   TI1++;
   }
   \If{env[i] <= threshold2 and env[i] > threshold1}{
   TI2++;
   }
   \If{env[i] <= threshold3 and env[i] > threshold2}{
   TI3++;
   }
   \If{env[i] <= threshold4 and env[i] > threshold3}{
   TI4++;
   }
   \If{env[i] > threshold4}{
   TI5++;
   }
   i++;
    }\;
q2= TI1+TI2;
    
    q3= TI3+TI4+TI5;
    
    \eIf{q2 >= q3} {goIdle=0;}{goIdle=1;}
   }{
   env[counter]=b;
   
   counter++;\;
  }
 \caption{Decide about system behavior based on a prediction of environmental behavior}
 \label{algorithmm}
\end{algorithm}

\begin{table*}[h]
\centering
\caption{Verification results}
\begin{tabular}{@{}ccc@{}}
\toprule
 & \textbf{Query} & \textbf{Result} \\ \midrule
1 & \textbf{E[<=10800; 20] (max:energy) } & 2899.54\\
2 & \textbf{E[<=10800; 20] (max:energy2) } & 2959.38\\
3 & \textbf{Pr[t<=10800](<>energy<=energy2) } & [0.901855,1]\\  
4 & \textbf{E<> Behavior.StandbyC and Behavior2.StandbyC and energy >= energy2} & Valid\\
5 & \textbf{A[] Behavior.StandbyC and Behavior2.StandbyC and energy <= energy2} & Invalid\\
\toprule
6 & \textbf{A[](Behavior.InIdle imply Behavior.angle1==-8 and Behavior.angle2==-79 and Behavior.angle3==-87 } & \\
 & \textbf{and Behavior.angle4==-105 and Behavior.angle5==87 and Behavior.angle6==-51)} & Valid\\
7 & \textbf{A[] (Behavior.ReadyA imply Behavior.angle1==-8 and Behavior.angle2==-87 and Behavior.angle3==-128 } & \\
 & \textbf{and Behavior.angle4==-56 and Behavior.angle5==87 and Behavior.angle6==-52)} & Valid\\
8 & \textbf{A[] (Behavior.StandbyC imply Behavior.angle1==83 and Behavior.angle2==-114 and Behavior.angle3==-95 } & \\
 & \textbf{and Behavior.angle4==-65 and Behavior.angle5==93 and Behavior.angle6==-49)} & Valid\\
9 & \textbf{A[] (Behavior.StandbyB imply Behavior.angle1==177 and Behavior.angle2==-123 and Behavior.angle3==-80 } & \\
 & \textbf{and Behavior.angle4==-66 and Behavior.angle5==90 and Behavior.angle6==-46)} & Valid\\
10 & \textbf{A[] (Behavior.BoxMovedC imply Behavior.e1<=15)} & Valid\\
11 & \textbf{A[] (Behavior.BoxMovedB imply Behavior.e1<=10)} & Valid\\

12 & \textbf{Behavior.BoxMovedC --> Behavior.Grab} & Valid\\
13 & \textbf{Behavior.BoxMovedB --> Behavior.Grab} & Valid\\
14 & \textbf{Behavior.InIdle --> Behavior.ReadyA} & Valid\\
15 & \textbf{Behavior.InIdle --> Behavior.Grab} & Valid\\
16 & \textbf{A[] not deadlock} & Valid\\
\bottomrule
\end{tabular}
\label{Query}
\end{table*}

As illustrated in Table \ref{energyconsumption}, the cost of going idle from position \textit{C}, and coming back to position \textit{A} is more than the cost of staying in standby mode for some time and then going to position \textit{A}. Therefore, it is beneficial to move to the idle mode when the duration of staying in that mode is longer than a certain amount of time. The proposed method examines whether such behavior is beneficial or detrimental to the system. Since the proposed algorithm makes decisions for each iteration (for every 20 boxes), and the decision is based on an average reduction in energy consumption, in some cases, the variable \textit{energy} may be more than \textit{energy2} (especially at the beginning due to small differences between energy and energy2), because by predicting the behavior of the environment, decisions are made for the entire next iteration. This is why query 5 (which indicates that energy consumption is always lower by having idle mode) is invalid. However, energy consumption generally decreases over time.   

The modeling and verification results (first five queries) show that, in the case of being aware of the environmental behavior, the system can optimize the energy consumption by improving the behavior of its components. The second group of verification results (queries 6 to 16) is related to the safety properties of the robotic arm. For example, in each of the different states, such as \textit{Idle}, \textit{A}, \textit{B}, and \textit{C}, the joints of the arm must be at a certain angle in order to achieve the ideal position.

\section{Related works} \label{Related works}

For optimizing the energy consumption, authors in \cite{paper9, paper10, paper13, paper14} have focused on behavioral techniques, such as time scaling, task scheduling, and break-release scheduling. 
In the time scaling technique, which refers to modifying energy consumption by scaling the task execution time, The optimization technique is not straightforward because the relation between energy consumption and task execution time is not linear.

Authors in \cite{solar}, presented a scheme for energy transfer between nodes. The authors used a solar panel with a certain efficiency and energy loss to charge the nodes. The solar panel was considered to be in clear-sky daily irradiation, which is almost ideal. Even issues such as energy loss can be increased or decreased under the influence of environmental conditions.

Authors in \cite{energyconstraints, Formalverification}, predict the energy consumption of the system and examine whether the energy consumption is in a certain range or not. In these works, the performance of the system is based on the analysis of environmental behavior. Both of these works could be improved by considering the behavior of the environment and its impact on energy consumption. Also, in \cite{security}, which examines the impact of security on system safety, the probability of a successful message is permanently fixed, something that may not always be true in reality.

\section{Discussion and future work} \label{future works}

Among the important challenges that CPS face is environmental behavior. The behavior of CPS is highly dependent on how the environment proceeds. In our case study, reducing energy consumption by changing the system's behavior depends on whether the boxes arrive earlier or later than a specific time. As mentioned in Section IV, the amount of energy consumption for the case study is reduced when the idle mode is not used in the case that the boxes' arrival time to the position \textit{A} is earlier than a specific time. Since the box arrival time is unpredictable, such environment behaviors cannot be decided by the system. This presents a challenge to the improvement of energy consumption. To solve this challenge, we dynamically modeled the environment's behavioral disorderliness and decided on the next iteration continuously.

In this work, we analyze the reduction of energy consumption by improving the behavior of the components without structural changes by using cost-optimal reachability analysis. Due to the dependence of system behavior on the environment, we examine the behavior of the environment from two aspects. First, the behavior of the environment may be quite irregular, so using fixed, predefined probabilities for incidents makes the existing system much simplified. For this purpose, to increase the system's flexibility against the disorderliness behavior of the environment, it is modeled dynamically. Second, the system can optimize its behavior by predicting the behavior that the environment may have in the future. So, by examining the past environmental behavior, the system can dynamically adjust its behavior to optimize its behavior in terms of energy consumption. We demonstrate the applicability of our framework on an industrial case study within the I4 Lab at SDU, which can also be extended to other assets in the CPS and IoT domains.

To increase system reliability, the system must be able to operate despite the behavioral disorder of the environment. This study demonstrates that it is possible to improve system performance by predicting its future behavior. Therefore, as future work, we intend to use machine learning (ML) in the decision-making process \cite{ML} to improve the way we predict environmental behavior and check the mathematical expectation of each incident. This helps the system make a decision, based on whether it will gain a profit or a loss by choosing a specific route. Moreover, by including the Entropy criterion \cite{Entropy}, we will measure the severity of this disorderliness as well as the efficiency of our decision-making algorithm in different conditions.

Since such systems communicate with the main server through IoT communication protocols, we will also include security checks. As a continuation of our previous work \cite{security}, which showed how a breach in cybersecurity could endanger safety, we will improve both the safety and security aspects of CPS with IoT features. As ongoing work, we will: (1) Use ML algorithms in the decision-making process to increase the efficiency of the decision; (2) Measure the severity level of the disorderliness of the environment by including the Entropy criterion; (3) Improve and integrate safety and security in CPS \& IoT verification; (4) Demonstrate the applicability with other verification/optimization tools supporting these techniques.    



\vspace{.1in}
\noindent \textbf{Acknowledgement}: This research has been  performed under the EF-CPS: Trustworthy CPS \& IoT within I4 project, and ``Energy-efficient Programming of Collaborative Robots” project funded by ELFORSK.

\bibliographystyle{IEEEtran}
\bibliography{fedcsis.bib}

\end{document}